\documentclass[prl,aps,twocolumn]{revtex4}
\usepackage{epsfig}
\usepackage{appendix}
\newcommand{\C}[1]{{\mathcal{#1}}}    

\begin{document}

\title{Effective Temperature Thermodynamics and the Glass Transition:\\ Connecting Time-Scales}
\author{Ido Regev$^{1,2}$, Xiangdong Ding$^{1}$ and Turab Lookman$^{1}$}
\affiliation{$^1$Theoretical Division Los Alamos National Laboratory, Los Alamos 87545, New Mexico\\
$^2$Center for Non-Linear Studies, Los Alamos National Laboratory, Los Alamos 87545, New Mexico}
\date{\today}
\begin{abstract}
We propose a theory based on simple physical arguments that describes a non equilibrium steady-state by a temperature-like parameter (an ``effective temperature''). We show how one can predict the effective temperature as a function of the temperature of the environment for a specific case of non-equilibrium behavior: radiation amorphization. The main idea that we present is that the amorphization process is inherently connected to the dynamical arrest that a liquid undergoes when it transforms into a glass. We suggest that similar arguments may hold also for the effective temperature under plastic deformation.
\end{abstract}

\maketitle
The idea that the state of a system in a non-equilibrium steady-state obeys a Gibbs-like distribution with a temperature-like parameter, called the ``effective temperature'' was studied extensively starting from the pioneering work of Edwards \cite{89Ed}. Edwards assumed that a granular material under external perturbation explores a set of equal probability states and therefore the micro-states are described by a stationary probability measure analogous to the Gibbs measure, but where a configurational temperature (that he called ``compactivity'') replaces the thermal bath temperature. Cugliandolo et al.\cite{97CKP} showed, for mean field models, that such an effective temperature has other properties desired for a thermodynamic description of non-equilibrium states, for example it controls the direction of heat flow. Furthermore, it was shown that fluctuation-dissipation theorems are violated in a very specific way: at short time scales the fluctuations are related to the susceptibilities in the usual way, described by the fluctuation dissipation theorem involving the thermal bath temperature. At large time-scales the usual fluctuation dissipation theorem  is not obeyed but satisfies an analogous expression that involves the effective temperature. In later work similar behavior was observed for sheared glasses at temperatures below the glass transition (about $66\%\, T_g$) \cite{02BB} and in sheared foam at zero temperature \cite{02OODLLR}. This suggests that a statistical mechanics description with a Gibbs-like measure might apply for these systems. It was also found that the response of amorphous materials to high shear-rates is related to their structure by an exponential function, which suggests that the structure might be described by a Gibbsian-like measure, similar to the ideas of Edwards \cite{07SKF,07HL}. Two important theories of plastic deformation in amorphous solids, the Shear Transformation Zone theory (STZ) \cite{98FL} and the Soft Glassy Rheology theory (SGR) \cite{97SLHC} use the idea of an effective temperature as a control parameter. In STZ theory it is assumed that the distribution of soft areas in a material obeys a Gibbs-like measure. Bouchbinder et al.\cite{07BLP} have extracted an effective temperature from non equilibrium values of an order parameter by inverting the equilibrium equation of state of the order parameter (assuming that the equation of state is still valid). In Bou\'e et al. \cite{10BHPRZ} this approach was taken one step further by showing explicitly that several coarse-grained structural properties of at least two different models of amorphous solids under extreme shear (plastic deformation) are described by a  Gibbsian measure controlled by an effective temperature. The value of the effective temperature as a function of the temperature was measured and was found to show non-monotonous behavior. It was also shown that the stress and the potential energy depend on the effective temperature. It is usually not possible to predict the effective temperature theoretically, which leads  to the use of fitting functions or scaling arguments. 

In this work we show, using molecular dynamics simulations, that the steady-state structure of an amorphous solid under irradiation can be described by a  Gibbs-like distribution. More importantly, we show how one can understand and {\it predict} the value of the effective temperature at the steady-state by connecting the fast and slow dynamics in the system in a way that is deeply rooted in the nature of the glass transition. Our explanations and theory are different than the ones given by Martin in  \cite{84M}, who predicted an infinite effective temperature at zero thermal bath temperature and also from the arguments given by Lund et al. in \cite{02LS} who suggested that the effective temperature has a finite value at zero bath temperature and from Bouchbinder et al. \cite{09BL} who suggested a thermodynamic explanation for the non-monotonic behavior of the effective temperature as a function of the thermal bath temperature.

{\bf The structure and slow dynamics of a supercooled liquid:} In a series of recent papers (cf. \cite{series,09BLPZ}) it was shown that the structural changes that a supercooled liquid undergoes on cooling can be represented by a finite set of species which are indexed by $1,2,\cdots, n$. The precise nature of these species may change from model to model, but they are always formed by particles and their nearest neighbors. The main advantage of these species is that they obey discrete statistical mechanics, in the sense that
their temperature-dependent concentrations $\langle C_i\rangle(T)$, are determined by a set of constant degeneracies $g_i$ and enthalpies $\C H_i$ such that:
\begin{equation}
\langle C_i\rangle(T) = \frac{g_i e^{-\C H_i/T }}{\sum_{i=1}^n g_i e^{-\C H_i/T}} \ . \label{CiT}
\end{equation}
After applying this description we always find that some species vanish at temperatures that are close to the glass transition temperature. This qualitative observation was made quantitative by summing the values of one or more of the concentrations of the vanishing species to what is called the `liquid-like' concentration $\langle C_\ell\rangle (T)$. The inverse of this concentration provides a length-scale (the typical distance between `fluid'  species):
\begin{equation}
\xi(T)\equiv [\langle C_\ell\rangle (T)]^{-1/d}\ , \quad \xi\to \infty ~{\rm when}~ T\to 0, \label{xi}
\end{equation}
where $d$ is the space dimension. It was amply demonstrated on a large variety of models that the
relaxation time $\tau_\alpha(T)$ measured using correlation functions in the super-cooled regime
is determined by this diverging scale according to
\begin{equation}
\tau_\alpha = \tau_0 e^{\mu \xi(T)/T} \ , \label{tauxi}
\end{equation}
where $\mu$ is a typical free energy per particle and $\tau_0$ a typical vibration time. In the following we will use the results of this analysis for a specific model: the ST (Shintani-Tanaka) two-dimensional liquid model \cite{06ST}. The ST model has $N$ identical particles of mass  $m$; each of the particles carries a unit vector ${\bf u}_i$ that can rotate on the unit circle. The particles interact via the potential $U(r_{ij} , \theta_i, \theta_j) = \bar U(r_{ij} ) + \Delta U(r_{ij} , \theta_i, \theta_j)$. Here $\bar U(r_{ij} )$ is the standard isotropic Lennard-Jones 12-6 potential, whereas the anisotropic part $\Delta U(r_{ij} , \theta_i, \theta_j)$ is chosen such as to favor local organization of the unit vectors of two interacting particles, such that they form $126^0$ with the radius-vector between them. This kind of interaction prefers structures with a five-fold symmetry, which frustrate crystallization. For full details of this model the reader is referred to Refs. \cite{06ST,07ILLP,09LPR}; here it suffices to know that with the parameters chosen in Ref. \cite{06ST} the model crystallizes upon cooling for $\Delta<0.6$ whereas for larger values of $\Delta$ the model exhibits all the standard features of the glass transition, including a spectacular slowing down of the decay of the correlation functions of the unit vectors $C_R (t) \equiv (1/N) \sum_i \langle {\bf u}_i(t)\cdot {\bf u}_i(0)\rangle$, which is very well described by Eq. (\ref{tauxi}). In \cite{07ILLP, 09LPR} and \cite{10BHPRZ} the above statistical mechanics theory was constructed for the ST model. We chose to separate the particles into three species, according to the possible number of neighbors interacting {\it in the most favorable way} with a given particle: particles with two favorable interactions are designated as belonging to the ``blue'' species ($n=2$), particles with one favorable interaction are labeled ``green'' ($n=1$) and ones with no favorable interactions at all are labeled ``red'' ($n=0$). In previous work we have found the exact numerical values of the constants $g_i$ and $\C H_i$ and showed that equation \ref{CiT} describes accurately the concentrations of species measured from experiments, and that equation \ref{tauxi} describes the typical relaxation times of the autocorrelation function $C_R (t)$ \cite{06ST, 07ILLP}.
\begin{figure}[htp]
\begin{center}
\includegraphics[width=0.4\textwidth]{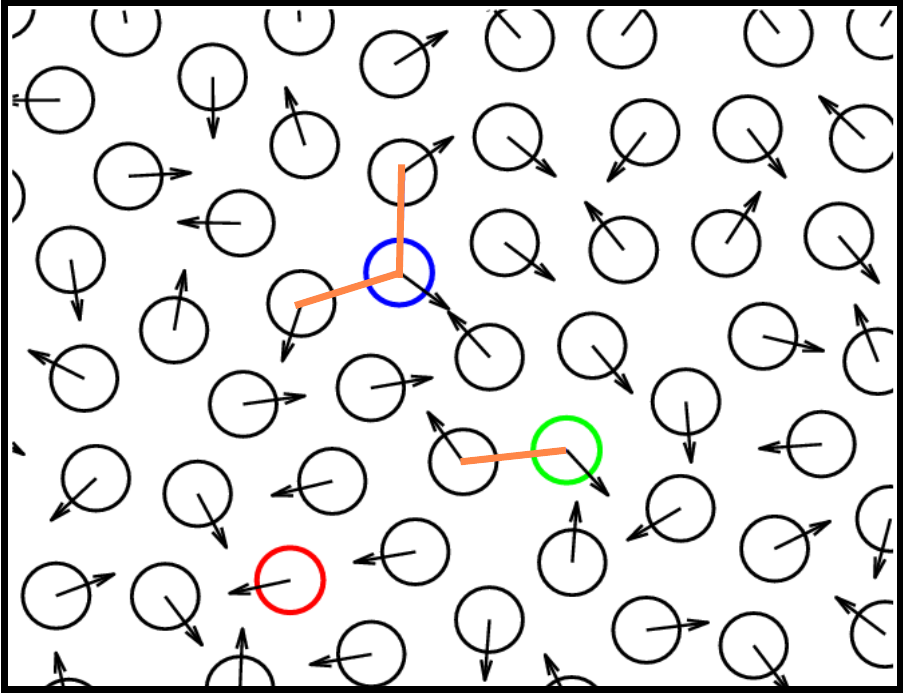}
\caption{(Color online): Representative configuration of particles in the ST model with examples for the three different species. Particle with two favorable interactions are assigned as a ``blue'' species, particles with one favorable interaction are labeled ``green'' and ones with no favorable interactions at all are labeled ``red''.}\label{fig1}
\end{center}
\end{figure}

{\bf Simulations:}
The molecular dynamics protocol we implemented was the following:  First, we carefully equilibrated a large number of independent configurations with $N=1024$ particles  in the NVT ensemble using the Berendsen thermostat over a wide range of temperatures. After this, we turned our equilibrium super--cooled liquids into amorphous solids by minimizing their potential energy using the conjugate gradient algorithm. This procedure can be thought of as quenching a liquid infinitely fast into a disordered solid whose temperature is formally $T=0$.  Then we irradiate at a constant rate while maintaining constant thermal bath temperature $T_b$. We ``irradiate'' the material by giving a particle a very high momentum at random. This simulates the effect of an energetic particle bombarding the material. The high momentum creates a cascade event where the particle transfers its energy to the surrounding particles. The simulations were performed at a constant temperature with a constant rate (average time between hits) and intensity (average momentum transfer) of radiation. The rate of bombardment of the material was much slower than the rate of thermal fluctuations and was chosen such that the material would not heat and melt.

\begin{figure}[htp]
\begin{center}
\includegraphics[width=0.45\textwidth]{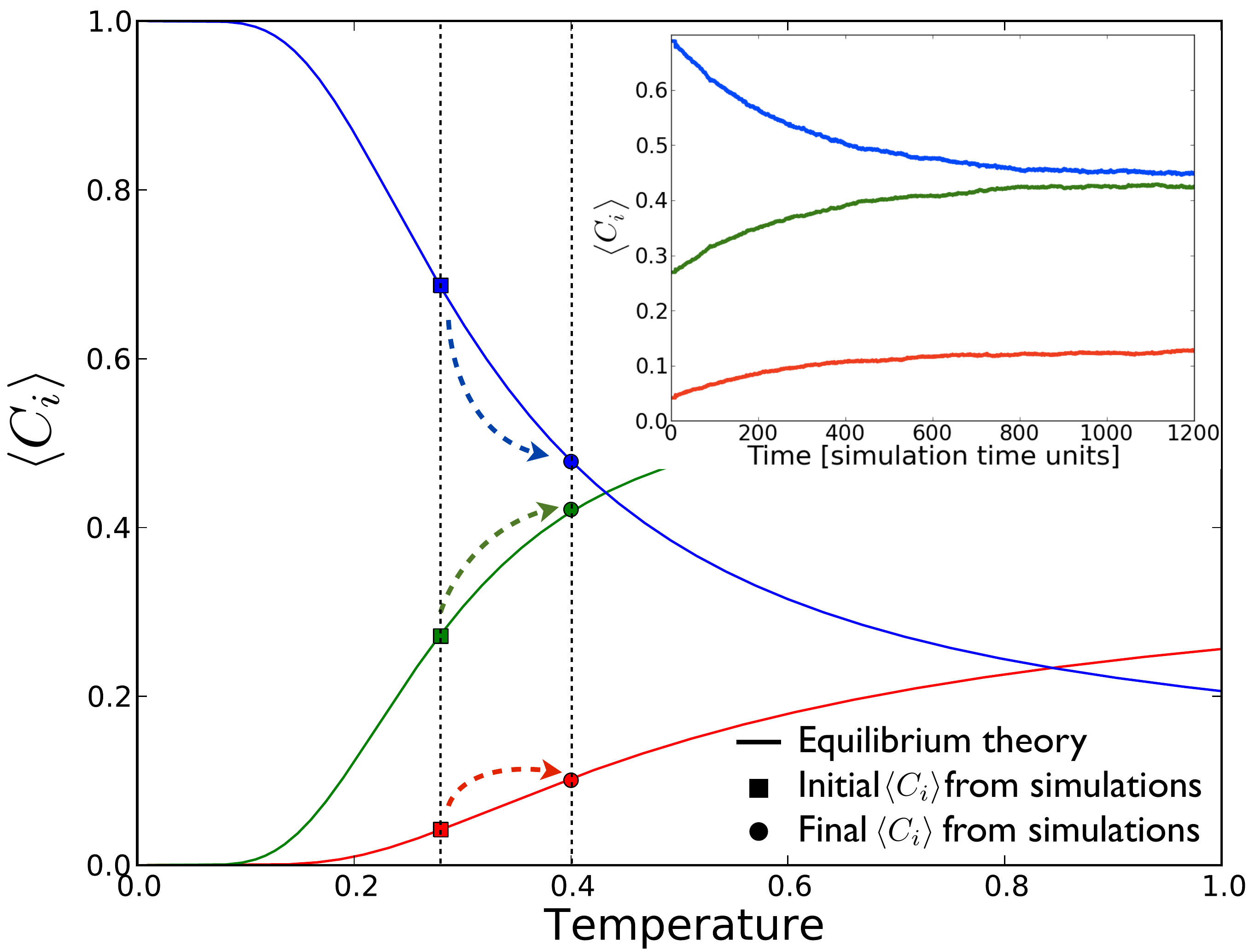}
\caption{(Color online): Concentrations of species with two strong bonds' (blue), one strong bond (green) and no bonds at all(red) as measured from simulations against the equilibrium theory (smooth lines, colors the same as concentrations). Squares represent the concentration at an initial random configurations quenched from a temperature $T=0.29$. Circles represent the concentrations of the species at non-equilibrium steady-state under radiation. Inset: concentrations as function of time starting from an amorphous solid.}\label{fig2}
\end{center}
\end{figure}

{\bf Effective temperature}: In \cite{10BHPRZ} we found that the concentrations of species in steady-state plastic deformation $\langle C_i\rangle_{ss}$ are described by the following Gibbsian distribution:
\begin{equation}
\langle C_i\rangle_{ss} = \frac{g_i e^{-\C H_i/ T_{eff} }}{\sum_{j=1}^n g_j e^{-\C H_j/ T_{eff}}} \ , \label{CiT_{eff}}
\end{equation}
where $g_i$ and $\C H_i$ have the same values as in the equilibrium Gibbs distribution and the effective temperature $T_{eff}$ is a parameter that depends on the external forcing and thermal bath temperature in a non-trivial way. In figure \ref{fig2} we show the concentrations $\langle C_i\rangle$ measured from simulations in an initial amorphous configuration (squares) and the concentrations measured at the steady-state under irradiation (circles). The curves are the theoretical concentrations given by the equilibrium Gibbs measure for this system - equation $\ref{CiT}$. The concentrations in the initial quenched state follow the Boltzmann distribution. This is an indication that the structural aspects that are reflected in the species are not sensitive to the quench. Under irradiation the concentrations undergo a significant change (see inset) and reach new steady-state values. Remarkably, the new concentrations, despite having changed significantly, still correspond to values of equations 1 with some ``effective'' temperature. For different radiation rates and thermal bath temperature the steady-state concentrations reach different values but they all agree well with the equilibrium Gibbs distribution with some effective temperature.

\begin{figure}[htp]
\begin{center}
\includegraphics[width=0.5\textwidth]{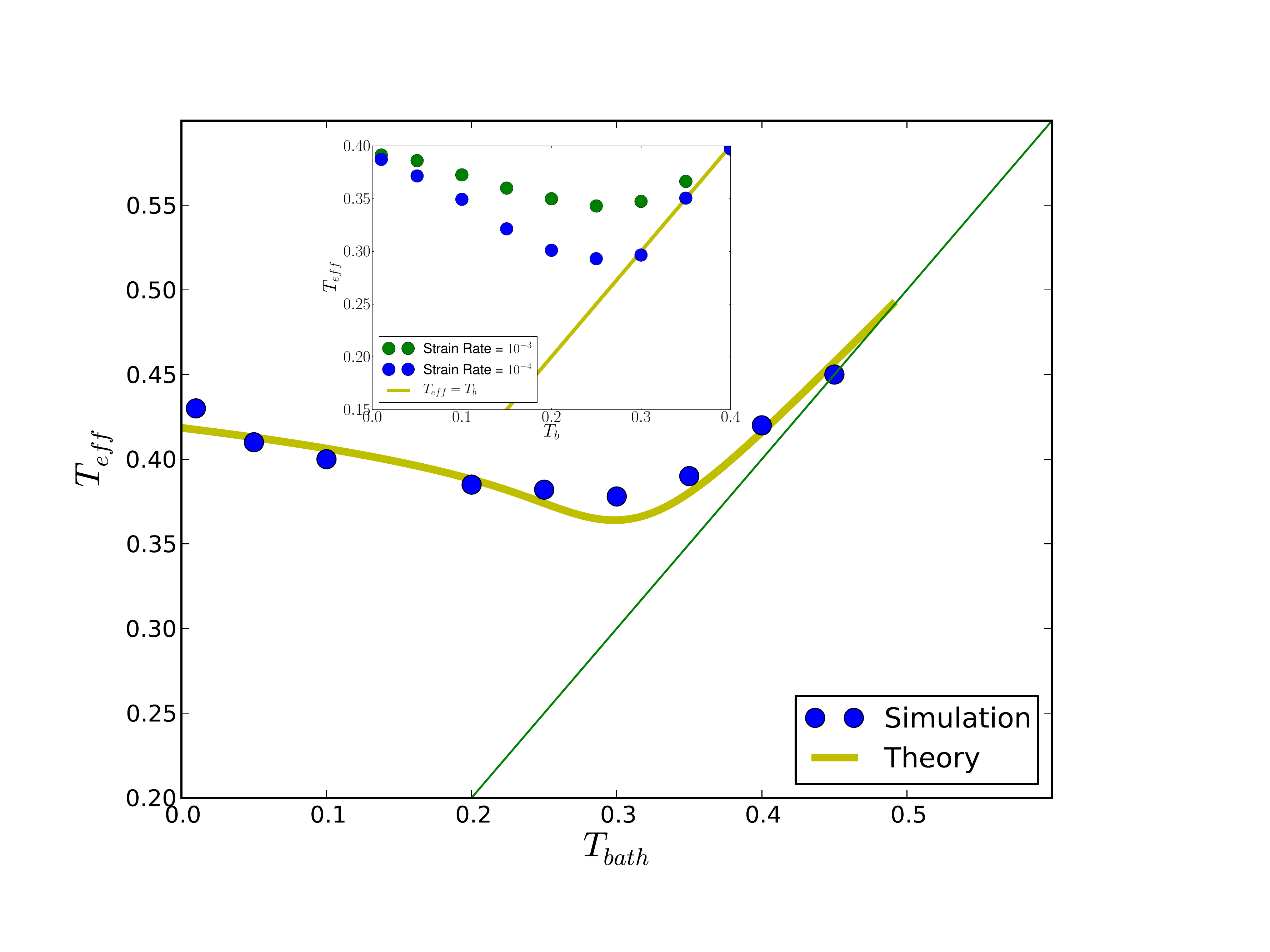}
\caption{(Color online): Effective temperature measured from simulations with the same irradiation power but different  thermal bath temperatures. The yellow line is the theoretical result. Inset: same result for simulations of plastic deformation with the same model at two different strain-rates.}
\label{fig3}
\end{center}
\end{figure}

{\bf Theoretical description of the steady-state effective temperature}:
In figure \ref{fig3} we show the effective temperature measured at the steady-state for a given radiation rate as a function of the thermal bath temperature (blue circles). In the inset we can see the steady-state values of the effective temperature of the same model under plastic deformation at two different, constant strain rates (see \cite{10BHPRZ}). One can observe that in both cases the behavior is quite similar: At high temperatures the dynamics is dominated by the fluctuations imposed by the thermal bath and the irradiation has little effect. In that regime the equilibrium Gibbs distribution accurately describes the statistics. As the temperature is lowered, the rate of irradiation (strain-rate in the case of plasticity) becomes comparable to the time scale associated with thermal fluctuations (eq. \ref{tauxi}). At that point the effective temperature starts to deviate from the thermal bath temperature. What happens at lower temperatures is completely un-intuitive: as the temperature is lowered, the effective temperature decreases moderately and than increases again until it reaches a finite effective temperature at zero temperature. But what is determining the values of the effective temperature? In the following we will give a physical explanation of this behavior. The basic idea underlying our analysis is that whenever an energetic particle transfers its energy to one of the particles in the material, it transfers some of its momentum to the neighboring particles, which in-turn transfer some of their momentum to their neighbors and so on, in a ``radiation cascade''. In many different materials, this is equivalent to a local melting of the material around the point of impact. Since the rest of the material is kept at a much lower temperature, the kinetic energy dissipates rapidly and the region cools down. Since the cooling process is very fast it will result in a {\it local glass transition}. We can describe this process using two rate equations (based on Fourier's cooling law):
\begin{eqnarray}
\dot{T}_{_K} &=& -\frac{1}{\tau_{th}}[T_{_K}-T_{_b}],\\
\dot{T}_{_{eff}} &=& -\frac{1}{\tau_{\alpha}(T_{_K})}[T_{_{eff}}-T_{_K}].\label{equations}
\end{eqnarray}
The first equation describes the relaxation of the kinetic temperature $T_{_K}$ of the molten region.  After long enough time, all of the kinetic energy will dissipate and the kinetic temperature will equilibrate with the thermal bath temperature $T_b$. The rate of relaxation is controlled by the thermal conductivity of the material and is material dependent. The second equation describes an aging process. At times longer than the relaxation time $\tau_{\alpha}(T)$ the structure of the molten area - described by the effective temperature $T_{eff}$ - reaches thermal equilibrium (note that one can use instead of equation \ref{tauxi}, the Vogel Fultcher expression to a similar effect). The solution to the first equation is:
\begin{equation}
T_{_K}(t,T_b) = [T_{_K}(t=0)-T_b]e^{-t/\tau_{th}} + T_b,
\end{equation}
and to the second equation:
\begin{equation}
T_{_{eff}}(t,T_b) = \frac{\int_0^t e^{\int_0^{t'} \frac{dt''}{\tau_{\alpha}[T_{_K}(t'',T_b)]}} \frac{T_{_K}(t',T_b)}{\tau_{\alpha}[T_{_K}(t',T_b)]} dt' + T_{_{eff}}(t=0)}{e^{\int_0^{t} \frac{dt'}{\tau_{\alpha}[T_{_K}(t',T_b)]}}}. \label{analytic}
\end{equation}
By choosing a cutoff time $\tau_{cf}$ we get an expression for the effective temperature as a function of the thermal bath temperature $T_{_{eff}}(T_b)=T_{_{eff}}(\tau_{cf},T_b)$. This equation is analytically solvable if we approximate and assume that $T_{_{K}}=T_b$. However, for small temperatures this assumption fails and the approximation is not accurate. The non-Arrhenius increase of the relaxation time at low temperatures means that the molten area will not reach the equilibrium structure but will undergo a glass transition and get ``stuck'' in some metastable state. In order to find the effective temperature that corresponds to this state, we numerically integrate the above equations with a high initial kinetic temperature which we guess to be: $T_{_K}(t=0)=T_{_{eff}}(t=0)=0.5$, representing the temperature of the molten region just after a cascade event. We also chose $\tau_{th}=200$ and introduce a time-cutoff $\tau_{cf}=400$ after which we assume that the structure does not evolve. The time scale $\tau_{cf}$  is related to the frequency of cascade events since this frequency controls the typical time between re-melting of a specific region. In figure \ref{fig3} we observe the agreement between a numerical solution of the equations and the data from simulations. The behavior of the effective temperature as a function of the thermal bath temperature is captured by the theory. This seems to apply, at least qualitatively, to plastic deformation. Despite being a very different process, plasticity in amorphous solids shares similarities with radiation damage: plastic events tend to be localized, a large amount of kinetic energy is released in the process from these localized regions, and the events result in irreversible structural changes.

{\bf Discussion}: We have constructed a theory for the values of the steady-state effective temperature of a material under irradiation. This result is based on a deep connection between the effective temperature and the glass transition which was proposed in the past \cite{10LBL} and is explained here. Due to the obvious similarity to the behavior of the effective temperature in plastic deformation, we suggest that a modified theory will be able to predict the effective temperature in plastic deformation as well. Future work will focus on applying the same formalism to more realistic simulations and experiments and also on relating structural properties to the effective temperature, for example in irradiation induced flow (see for example \cite{03MAAA}). 

\acknowledgments
Ido Regev would like to thank Laurent Bou\'e for very useful discussions.
This work was supported by the US Department of Energy's Los Alamos National Laboratory through contract DE-AC52-06NA25396.

\end{document}